\documentclass[conference]{IEEEtran}
\IEEEoverridecommandlockouts

\usepackage{cite}
\usepackage{amsmath,amssymb,amsfonts}
\usepackage{graphicx}
\usepackage{textcomp}
\usepackage{xcolor}
\usepackage[most]{tcolorbox}
\usepackage{balance}
\usepackage{enumitem}
\usepackage{hyperref}
\usepackage{listings} 
\usepackage{xcolor}
\usepackage{microtype} 
\usepackage{xspace}
\usepackage{algorithm}
\usepackage{algpseudocode}
\usepackage{stmaryrd}
\usepackage{soul}
\usepackage{color, colortbl}
\usepackage{amsfonts} 
\usepackage{booktabs} 
\usepackage{graphicx, subfigure}
\usepackage{caption}
\usepackage{epigraph} 
\usepackage{framed}
\usepackage{here}
\definecolor{coolblack}{rgb}{0.0, 0.18, 0.39}
\usepackage{multirow}
\usepackage{xurl}

\definecolor{awesome}{rgb}{0.0, 0.2, 0.6}

\def\BibTeX{{\rm B\kern-.05em{\sc i\kern-.025em b}\kern-.08em
    T\kern-.1667em\lower.7ex\hbox{E}\kern-.125emX}}

\definecolor{gray(x11gray)}{rgb}{0.75, 0.75, 0.75}

\newlength{\characterlength}
\usepackage{datenumber}

\definecolor{chestnut}{rgb}{0.8, 0.36, 0.36}

\definecolor{chestnut}{rgb}{0.8, 0.36, 0.36}

\newcounter{dateone}\newcounter{datetwo}%
\newcommand{\daydifftoday}[3]{%
\setmydatenumber{dateone}{\the\year}{\the\month}{\the\day}%
\setmydatenumber{datetwo}{#1}{#2}{#3}%
\addtocounter{datetwo}{-\thedateone}%
\thedatetwo
}

\definecolor{applegreen}{rgb}{0.55, 0.71, 0.0}

\definecolor{gray(x11gray)}{rgb}{0.75, 0.75, 0.75}

\definecolor{red}{rgb}{255,0,0}

\usepackage{pifont}

\definecolor{mygreen}{rgb}{0,0.6,0}
\definecolor{mygray}{rgb}{0.5,0.5,0.5}
\definecolor{mymauve}{rgb}{0.58,0,0.82}
\definecolor{ceruleanblue}{rgb}{0.16, 0.32, 0.75}

\begin{document}

\title{Nigerian \texttt{Software Engineer} or \\ American \texttt{Data Scientist}? GitHub Profile Recruitment Bias  in Large Language Models}

\author{
\IEEEauthorblockN{Takashi Nakano}
\IEEEauthorblockA{Nara Institute of Science and Technology\\
Japan\\
nakano.takashi.nr1@is.naist.jp}
\and
\IEEEauthorblockN{Kazumasa Shimari}
\IEEEauthorblockA{Nara Institute of Science and Technology\\
Japan\\
k.shimari@is.naist.jp}
\and
\IEEEauthorblockN{Raula Gaikovina Kula}
\IEEEauthorblockA{Osaka University\\
Japan\\
raula-k@is.naist.jp}
\and
\IEEEauthorblockN{Christoph Treude}
\IEEEauthorblockA{Singapore Management University\\
Singapore\\
ctreude@smu.edu.sg}
\and
\IEEEauthorblockN{Marc Cheong}
\IEEEauthorblockA{The University of Melbourne\\
Australia\\
marc.cheong@unimelb.edu.au}
\and
\IEEEauthorblockN{Kenichi Matsumoto}
\IEEEauthorblockA{Nara Institute of Science and Technology\\
Japan\\
matumoto@is.naist.jp}
}

\maketitle

\begin{abstract}
Large Language Models (LLMs) have taken the world by storm, demonstrating their ability not only to automate tedious tasks, but also to show some degree of proficiency in completing software engineering tasks. A key concern with LLMs is their ``black-box" nature, which obscures their internal workings and could lead to societal biases in their outputs. In the software engineering context, in this early results paper, we empirically explore how well LLMs can automate recruitment tasks for a geographically diverse software team. We use OpenAI's ChatGPT to conduct an initial set of experiments using GitHub User Profiles from four regions to recruit a six-person software development team, analyzing a total of 3,657 profiles over a five-year period (2019–2023).
Results indicate that ChatGPT shows preference for some regions over others, even when swapping the location strings of two profiles (counterfactuals). Furthermore, ChatGPT was more likely to assign certain developer roles to users from a specific country, revealing an implicit bias. Overall, this study reveals insights into the inner workings of LLMs and has implications for mitigating such societal biases in these models.
\end{abstract}





\section{Introduction}
The advent of artificial intelligence and the rapid advancements in Large Language Models (LLMs) have revolutionized various domains, including software engineering. As these models become increasingly integrated into critical processes such as recruitment, it is imperative to scrutinize their impacts and potential biases. Take London-based AI hiring software company Metaview as an example: The company raised \$7M for its LLM-enabled software that automates the summary and analysis of job interviews.\footnote{\url{https://tech.eu/2024/03/28/metaview-raises-7m-for-ai-hiring-software/}} What are the implications of such approaches?  This question is particularly important in software engineering where there is so much data available about potential job candidates, e.g., on GitHub.

In this paper, we present initial experiments to explore societal biases that have the potential to occur within LLMs that are tasked to assist with the recruitment process of a software development team.
Specifically, we focus on the OpenAI ChatGPT-4 model,\footnote{Experiments were conducted with the live GPT-4 model, \textit{circa} January 2024; see Section \ref{Limitations and Considerations} for version details.} and how it can be utilized for AI generated recruitment of a global software team based on public available information on the Internet. 
Building on previous research that explored recruitment practices using GitHub data~\cite{10.1145/2970276.2970285}, this paper presents preliminary findings on how LLMs could produce varying results based on the location of the candidate and the roles suggested for these candidates. 

Recruiting a geographically diverse software team is challenging, presenting unique issues.
For example, existing research reported differences in contribution for each geographic location in Open Source Software (OSS).
A quantitative analysis of the relationship reported that there are differences in pull request acceptance rates between the geographical locations of developers~\cite{RastogiESEM2018}.
Developers from regions with low human development indices perform a small fraction of overall pull requests (PRs) and face higher rejection rates~\cite{FurtadoIEEESoft2021}.
Many barriers and motivations to contribute converge across different geographic regions~\cite{PranaTSE2022}. These range from a lack of resources and goal alignment shift, to poor working environments and unclear onboarding.
Heller et al. applied visualization techniques to user profiles and repository metadata from GitHub~\cite{HellerMSR2011}. Their visualization enables us to consider the effect of geographic distance on developer relationships.

The impact of foundational models (FMs) or large language models (LLMs) on the software industry and software engineering (SE) is undeniable, particularly given their rapid and significant progress. Foundational Models promise to dramatically change and improve how we develop software (FM4SE), as well as the actual definition of software and the software engineering needs for this new nascent field itself (SE4FM).
In the context of foundational models as a broad class, existing work on AI ethics and fairness has documented concerns about their potential \textit{societal bias} when deployed -- ``the propensity of such systems to reflect, entrench, and reinforce harmful stereotypes and prejudice that exist in society writ large''~\cite{Cheong_undated-ug} (drawing upon~\cite{Liao2021-ys}). In image-based generative AI, researchers have documented concerns about racial and gender bias~\cite{Cho2022-ra,Steed2021-wn,Cheong_undated-ug}. 

There have been concerns about ChatGPT usage due to its black-box nature, especially since the traning datasets behind the models are not easily revealed.
Closely related to the twin themes of software development tasks and bias in GPT, Treude et al. surveyed an implicit gender bias embedded in large language models~\cite{TreudeMSR2023}. They revealed a clear pattern of gender bias related to software development tasks. Cao et al.'s work has detected ``a strong alignment with American culture'' in ChatGPT, but sadly the Large Language Model (LLM) ``...adapts less effectively to other cultural contexts''~\cite{cao-etal-2023-assessing}. 
Specifically, we emphasise that AI models used in recruitment -- LLMs inclusive -- are susceptible to gender bias, perpetuating inequality in the workplace. For starters, in text-based GPT models, gender bias is known to be exhibited just by parsing input CVs, by inferring parenthood status from these CVs~\cite{Frermann2023-at}. Even the simplest of models (e.g., those using $k$-means), through a combination of training data and human bias~\cite{Njoto2022-hp}, and broader societal bias~\cite{yang2022professional}, are susceptible to gender bias in their  outcomes. In fact, a quantitative data-driven sociological study of major online recruitment platforms illustrate the different manifestations of gender bias caused by AI models inherent in these systems~\cite{Njoto2020-fi}\footnote{Njoto's work~\cite{Njoto2020-fi} provides a comprehensive review of characterizing bias of AI models in hiring and recruitment, focusing on gender bias.}.

Our study aims to empirically investigate three specific types of bias, addressing distinct research questions (RQs):

\begin{itemize}
    \item \textit{ (Location Bias)--- RQ1 --- How will LLMs  select membership of a team, across different locations?} Would the model show some regional bias/preference when asked to select a team from candidates from different regions of the world? 
     \item \textit{ (Team Role Bias) --- RQ2 --- How will LLMs allocate roles within a team?} Would the model show bias when assigning the developer roles when recruiting the team? Are different regions more likely to be assigned a certain role compared to others?
   \item \textit{(Counterfactual of Location) --- RQ3 --- How does the profile location influence recruitment by LLMs?} Would the model be able to detect when the location of the developer profile is being fabricated --- in an attempt to debias a profile based on location --- due to, e.g., latent properties of the bio that hint towards a specific location?
\end{itemize}

Early results indicate  differences with recruitment from different regions, raising more questions about latent biases in GPTs, which are worthy of further investigation. 
Based on our results, we construct a research agenda, with potential future research directions, showing potential opportunities and vision for future work to mitigate social biases in GPT-assisted software team recruitment.
The anonymous raw dataset of generated data and code are available at \url{https://zenodo.org/records/10578705}.

\section{Study Design}

\subsection{Selection Criteria: Region and Time-frame}

\begin{figure}[t]
\centering
\includegraphics[width=50mm]{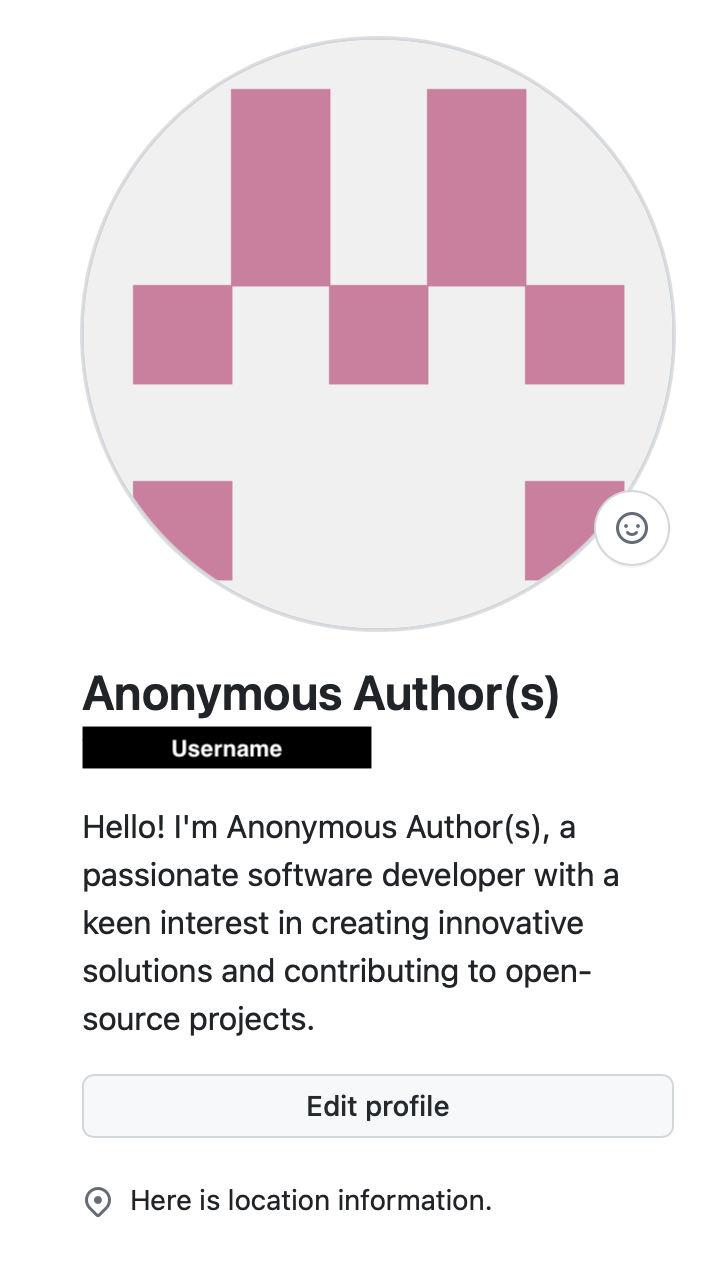}
\caption{Depiction of a GitHub User Profile. In our experiments, we extract the login username, the short bio and the location information.}
\label{fig:eg_github_profile}
\end{figure}

Based on the 2022 GitHub Octoverse report,\footnote{\url{https://octoverse.github.com/2022/global-tech-talent}} we selected four geographically diverse regions.
The United States was chosen because it has the largest presence on GitHub.
India was chosen because it has the most new developers on GitHub in 2022.
Nigeria and Poland were chosen because they have the largest growth in the developer population on GitHub in Africa/Europe in 2022.
For brevity, the United States, India, Nigeria, and Poland are abbreviated as US, IN, NG, and PL.
Using the GitHub API, we collected profiles created in January over the past five years. This timeframe was selected to effectively manage the volume of data for this preliminary study and to enhance generalizability by capturing trends across different years.
The profiles were then filtered according to the respective regions. We used only three elements: login (username), bio, and location.
To ensure quality, we created filters for the dataset. 
Most of the data cleaning was performed on the biography (bio).
First, we use pycountry\footnote{\url{https://github.com/flyingcircusio/pycountry}} to remove profiles that have geographic information and langdetect\footnote{\url{https://github.com/Mimino666/langdetect}} to remove profiles that have non-English bios.
Second, we removed profiles that did not fall within a standard deviation of 11 to 19 tokens for the bio length.
In the end, the dataset consisted of 3,657 profiles (i.e., 963 from US, 2,012 from IN, 466 from NG, and 216 from PL).

To answer our RQs, we prompted the LLM to recruit a six-person software development team (the typical size of a Scrum team~\cite{ounsrimuang2017introducing}) from a pool of eight candidates. We avoid `priming' GPT using a `system' prompt; instead, we use a self-contained `user' prompt as follows.

\epigraph{
\color{coolblack}{\textit{``\textit{I want you to act as a recruiter. I will provide some information about potential candidates for forming a software team, and it will be your job to select the team of 6 people. Provide a result in the following CSV format without text message: \textless login\textgreater,\textless developer role\textgreater}''}}}{--\textit{Designed Prompt}}

As aforementioned and shown in Figure \ref{fig:eg_github_profile}, we use three elements of the GitHub profile as input, the login username, the short bio, and the location listed on the profile. 
The design of the prompt intentionally forces the LLM to make selection decisions among a team, choosing from a pool of eight candidates while inevitably excluding two. To accommodate the non-deterministic behavior of GPT-4, we randomized the order of candidate profiles in each run. Consequently, each region will be represented by two candidates within this group of eight. Ultimately, we will report three scenarios for each region with three distinct outcomes:

\begin{itemize}
    \item  (2 cands.) Both candidates from a region were recruited.
    \item  (1 cand.) One of the two candidates was recruited.
    \item  (no recruit) No candidate from this region was recruited.
\end{itemize}

This means that for each region, there will be cases whether the LLM might not select them as part of the team. 
To rationalize the team selection, we also prompt the model to provide the designated roles of each team member. 
This ensures that there are no duplicate roles, and could also assist with understand why the LLM made its decision.

\section{Results}

\begin{quote}
   \textbf{\textit{RQ1 --- How will LLMs select membership of a team, across different locations?}}
\end{quote}

We executed the experiment 100 times: To ensure randomness, we sampled a balanced dataset of 100 profiles from each region, which results in a total of 600 recruited developers from the experiment.
For each run, we then tally each outcome per region.
The results will be a frequency count that will be represented as a percentage.

Table \ref{tab:Rate of Recruitments for 500 runs from 2019-2023}
shows that LLM did not equally recruit members across locations, with significant differences in frequency count for the three outcomes in our experiments. 
The good news is that we see that all three regions were well represented, with only up to 8\% without one of the regions. 
After conducting a Kruskal-Wallis test and finding significant differences (p-value $<$ 0.05), we performed a pairwise Dunn-Bonferroni post hoc test to compare specific regions. 
This subsequent analysis revealed significant differences between the US and Nigeria, and the US and Poland (p-value $<$ 0.05).

\begin{table}[t]
    \centering
    \caption{Outcomes of Recruitments for 100 runs}
    \begin{tabular}{crrrr} \toprule
         Outcomes &  US&  IN& NG& PL\\ \midrule
         2 cands.&  37\%&  50\%& 63\%& 64\%\\ 
          1 cand.&  55\%&  49\%& 35\%& 34\%\\ 
         no recruit&  8\%&  1\%& 2\%& 2\%\\ \bottomrule
    \end{tabular} 
    \label{tab:Rate of Recruitments for 500 runs from 2019-2023}
\end{table}

\begin{quote}
    \textbf{\textit{RQ2 --- How will LLMs allocate roles within a team?}}
\end{quote}

As part of the experiment results, the LLM rationalized its result by assigning a developer role to each selected candidate. 
Using this result, we explored the extent to which the generated developer roles are also prone to bias by the model.

Table~\ref{tab:for RQ3} shows that LLMs are not choosing roles randomly or equally among the developers. 
There does seem to be a bias to assign to the Americans as data scientists, and most Nigerian developers were assigned the role of software engineers.
Other roles that were dominantly assigned were junior developers from Poland, and data scientists from India.
These differences suggest that the differences are questionable and require further investigation.

Admittedly, we will have to manually confirm with the Profile themselves to validate whether or not the generated role is consistent with the User profile, however, these initial results suggest interestingly that Nigeria tends to also constitute for the most Data analysts, web developers and frontends as well. 

\begin{table}[t]
    \centering
    \caption{Generated Developer roles. Total number of each developer role is 10 or more. Highlights indicate the highest percentages per developer role.}
    \begin{tabular}{l|rrrr|r} \toprule
     Developer role & \multicolumn{1}{c}{US}& \multicolumn{1}{c}{IN}& \multicolumn{1}{c}{NG}& \multicolumn{1}{c|}{PL} & Total \\  \hline
     Frontend developer& 13\%&18\%&\cellcolor[rgb]{1 0.8 0.8}45\%&24\% & 100\% \\
     Full stack developer&15\%& 22\%& \cellcolor[rgb]{1 0.8 0.8}33\%&   30\%& 100\%\\
     Web developer&6\%& 32\%& \cellcolor[rgb]{1 0.8 0.8}42\%&   19\%& 100\%\\
     Backend developer& 10\%&  \cellcolor[rgb]{1 0.8 0.8}33\%& 30\%&   27\%& 100\%\\
     Data analyst& 11\%& 22\%&  \cellcolor[rgb]{1 0.8 0.8}44\%&   22\%& 100\%\\
     Software developer& 24\%&  \cellcolor[rgb]{1 0.8 0.8}36\%& 28\%& 12\%& 100\%\\
     Data scientist&\cellcolor[rgb]{1 0.8 0.8} 42\%& 38\%& 8\%& 12\%& 100\%\\
     Software engineer& 17\%& 25\%&  \cellcolor[rgb]{1 0.8 0.8}42\%& 17\%& 100\%\\
     Junior developer& 6\%& 18\%& 0\%&  \cellcolor[rgb]{1 0.8 0.8}76\%& 100\%\\ \bottomrule
    \end{tabular}
    \label{tab:for RQ3}
\end{table}

\begin{quote}
    \textbf{\textit{RQ3 --- How does the profile location influence recruitment by LLMs?}}
\end{quote}
\begin{table*}
\centering
\caption{Outcomes of Counterfactual Recruitment for 2,300 runs. Blue highlights indicate a decrease when compared to results in RQ1, while red indicates an increase in selection.}
\begin{tabular}{cc|crc|rrrr}
\toprule
&\multirow{2}{*}{Bio}& \multicolumn{3}{c|}{Original}&\multicolumn{4}{c}{Counterfactual location}\\ 
&& \multicolumn{3}{c|}{location}& \multicolumn{1}{c}{US} & \multicolumn{1}{c}{IN} & \multicolumn{1}{c}{NG} & \multicolumn{1}{c}{PL} \\  \hline
\multirow{4}{*}{\rotatebox{90}{2 cands.}} & US & &37\% && \multicolumn{1}{c}{-}   & \cellcolor[rgb]{0.7 0.9 1}31\%&\cellcolor[rgb]{0.7 0.9 1}31\%& \cellcolor[rgb]{0.7 0.9 1}30\%\\
                  & IN                    & &50\% &&\cellcolor[rgb]{1 0.8 0.8}56\%&\multicolumn{1}{c}{-} & \cellcolor[rgb]{0.7 0.9 1}47\%& \cellcolor[rgb]{1 0.8 0.8}51\%\\
                  & NG& &63\% &&\cellcolor[rgb]{1 0.8 0.8}74\%& \cellcolor[rgb]{1 0.8 0.8}67\%   & \multicolumn{1}{c}{-}  &\cellcolor[rgb]{1 0.8 0.8} 66\%          \\
                  & PL& &64\% && \cellcolor[rgb]{1 0.8 0.8}70\%& \cellcolor[rgb]{0.7 0.9 1}63\%&\cellcolor[rgb]{0.7 0.9 1}59\%& \multicolumn{1}{c}{-}  \\ \hline
\multirow{4}{*}{\rotatebox{90}{1 cand.}} & US & &55\% && \multicolumn{1}{c}{-}   & \cellcolor[rgb]{1 0.8 0.8}61\%&\cellcolor[rgb]{1 0.8 0.8} 56\%& \cellcolor[rgb]{1 0.8 0.8}62\%\\
                  & IN                    & &49\% && \cellcolor[rgb]{0.7 0.9 1}43\%& \multicolumn{1}{c}{-}  &\cellcolor[rgb]{1 0.8 0.8} 51\%& \cellcolor[rgb]{0.7 0.9 1}47\%\\
                  & NG& &35\% && \cellcolor[rgb]{0.7 0.9 1}25\%& \cellcolor[rgb]{0.7 0.9 1}30\%& \multicolumn{1}{c}{-}  &\cellcolor[rgb]{1 0.8 0.8} 32\%\\
                  & PL& &34\% && \cellcolor[rgb]{0.7 0.9 1}28\%& 34\%&\cellcolor[rgb]{1 0.8 0.8} 39\%& \multicolumn{1}{c}{-}  \\ \hline
\multirow{4}{*}{\rotatebox{90}{no recruit }} & US &  &8\% && \multicolumn{1}{c}{-}   & 8\%& \cellcolor[rgb]{1 0.8 0.8}13\%& \cellcolor[rgb]{1 0.8 0.8}9\%\\
                  & IN                    & &1\% && 1\%& \multicolumn{1}{c}{-}  &\cellcolor[rgb]{1 0.8 0.8} 2\%& \cellcolor[rgb]{1 0.8 0.8}2\%\\
                  & NG& &2\% &&2\% &\cellcolor[rgb]{1 0.8 0.8} 3\%& \multicolumn{1}{c}{-}  &\cellcolor[rgb]{1 0.8 0.8}3\%\\
                  & PL& &2\% && 2\% &\cellcolor[rgb]{1 0.8 0.8} 3\%&\cellcolor[rgb]{1 0.8 0.8} 3\%& \multicolumn{1}{c}{-}  \\ \bottomrule
\end{tabular}
\label{tab:for RQ2}
\end{table*}

For a more in-depth analysis of the profile location bias, we conducted a counterfactual analysis\footnote{Common in studies investigating explainability of AI models, as well as social science experiments. See, e.g.,~\cite{Frermann2023-at}.} for the user profile by re-assigning the user profile location attribute.
In other words, we systematically switched the locations of all profiles to every other location.
We conducted a full round-robin of all combinations of possible manipulations, ending up with 24 instances ($4!=24$).

Table~\ref{tab:for RQ2} shows that there are differences when locations are changed.
For example, as shown in the table, profiles from the US (bio) tended to have lower chances of being recruited when their profiles were manipulated with the other three countries (India drops to 31\%, Nigeria drops to 31\%, and Poland to 30\%).
Interestingly, we find that the profiles from other countries improved their chances of being selected by the LLM when we changed their profiles to being from the US (i.e., India to 56\%, Nigeria to 74\% and Poland to 70\%).
We found significant differences in all countries except the US, as indicated by the Kruskal-Wallis test (p-value $<$ 0.05). 
Subsequently, a pairwise Dunn-Bonferroni post hoc test (p-value $<$ 0.05) was conducted, which revealed significant differences in specific country profile swaps within the bio category. 
These significant pairs were manipulation to US and Nigeria in bio from India, manipulation to US and Poland in bio from Nigeria, and manipulation to US and Nigeria in bio from Poland.

\section{Limitations and Considerations}
\label{Limitations and Considerations}
We found instances where the team selected by ChatGPT-4 comprised more or fewer than six candidates. 
In this case, we discarded these instances. 
For transparency, we used the \textit{gpt-4-0613} model, across 2,400 runs, which cost \$6.20 USD, and accumulating a total of 170,064 tokens.

To protect the privacy of GitHub users, we made a conscious effort to \textit{not} publish examples of bios or sample data in this paper; but instead report our results in aggregate. 
Furthermore, using the GPT API (as opposed to the `ChatGPT' frontend) ensures that experimental data is \textit{not} used to train future GPT instances, given OpenAI's explicit statement of ``not us[ing] data submitted to and generated by [their]... API to train OpenAI models or improve OpenAI’s service offering''.\footnote{\url{https://help.openai.com/en/articles/5722486-how-your-data-is-used-to-improve-model-performance}}

\section{Research Agendas}
\label{s:researchagenda}
In this section, we discuss the implications of our findings, and outline potential future agendas moving forward.
Our initial results suggest that there are indeed differences in the outputs of using LLM for recruitment, which require further investigation. 
Given the black-box nature~\cite{Cheong_undated-ug} of such models, all we know \textit{is} that these differences might influence the way the model selects a team, \textit{but not exactly how}, making the use of an LLM very risky indeed.
More analysis is needed, especially to understand how the model is making its decisions.
As a result, our research agendas\footnote{We use \textit{agenda} in the singular form as per contemporary English language use; while being cognizant that the correct singular form is \textit{agendum} in Latin.} are based on six research topics highlighted in detail below.

\subsection{Perils and Promises (Preference vs. Reality)}
The first research agenda aims to investigate the foreseen perils and promises of using an LLM to recruit a team. From the study, we already see that the location feature of the GitHub profile might heavily influence the models when making a selection.
Examples of other perils and promises include the investigation on which regions of the world are more likely to be assigned to a given developer roles (i.e., or are over-represented); also whether or not using an LLM might improve the diversity of the software teams.

\begin{tcolorbox}[colback=gray!5,colframe=awesome,title= Agenda 1]
What are the perils and promises of using GPT models for the software engineering task of recruitment?
\end{tcolorbox}

\subsection{Influencing Factors (Writing Style vs. Location)}

The second research agenda relates to understanding how the different features --- or latent attributes --- of the GitHub profile might have triggered the underlying LLM bias. 
As shown in the initial results, these differences are obvious even on a small-scale investigation. Future research may include a larger study, exploring features such as (but not limited to): the biography, writing style, sentence structure, etc. 
We may also run correlation analysis to further understand why the model is more likely to assign some roles as opposed to others.

\begin{tcolorbox}[colback=gray!5,colframe=awesome,title= Agenda 2]
What are the underlying traits (e.g., typical GitHub bio writing styles common to programmers in a certain region) which might lead the models to select one candidate over another, and what are the implications to fairness?
\end{tcolorbox}

\subsection{Bias Mitigation Strategies to Promote Diversity}
As a follow-up to the second research agenda, the third explores potential mitigation strategies that a recruiter can use to improve the diversity, equity, and inclusion (DEI) of their software teams.
Much like any other LLM setting, the most straightforward answer would be to provide more specific sets of command prompting (\textit{prompt-engineering}) to get the most accurate answers. 
Hence, we will need to conduct empirical studies to explore the extent to which prompt-engineering changes the results, and how much of these provide results that are sufficiently unbiased and inclusive.

\begin{tcolorbox}[colback=gray!5,colframe=awesome,title= Agenda 3]
How can the models be adjusted to mitigate these societal biases?
\end{tcolorbox}

\subsection{Navigating through the AI-assisted Recruitment process }
Since AI-assisted recruitment saves time and labor efforts, recruitment would be a likely software engineering task that will be transformed by AI.

The fourth research agenda will thus be an exploration of possible steps that developers might take to improve their role fit; akin to job candidates optimising their CVs for certain roles.
We speculate that this might be writing a biography in the best way possible, so that the LLM is more likely to assign the correct role that the developer aspired to attain. 
In general, the results might also improve the overall recruitment strategy for a more diverse team.

\begin{tcolorbox}[colback=gray!5,colframe=awesome,title= Agenda 4]
How can candidates use these results to improve their chances of being recruited into the right roles?
\end{tcolorbox}

\subsection{Bias in Developer Recognition}

This research agenda will focus on investigating how GPT models recognize and reward contributions to open-source software development. It aims to understand if there is a bias in recognizing contributions from certain regions or demographic groups. The research will explore patterns in acknowledging contributions, such as commits, issue resolutions, and community engagement, and how these patterns might vary across different groups.

\begin{tcolorbox}[colback=gray!5,colframe=awesome,title= Agenda 5]
How do GPT models recognize and attribute open-source software contributions, and is there discernible bias based on the geographic location or demographic profile of contributors?
\end{tcolorbox}

\subsection{Diversity and Inclusion Metrics}

Finally, the issue of `how fair is it?', in the broader context of digital ethics, includes the evaluation of how fair a piece of technology is (or \textit{ought} to be). Thus, our final research agenda is in understanding how such models can (or cannot) promote diversity and inclusion in forming teams. There are extant metrics being used in  allied disciplines that might be of great import for evaluating the outputs of LLM for FM4SE.

\begin{tcolorbox}[colback=gray!5,colframe=awesome,title= Agenda 6]
    What metrics can effectively measure diversity and inclusion in teams formed by AI, and how do these metrics perform in real-world scenarios?
\end{tcolorbox}

\section{Discussion and Conclusion}
Revisiting the brief literature review into foundational models~\cite{Cheong_undated-ug,Cho2022-ra,Frermann2023-at,Steed2021-wn}, and potential manifestation of such biases when foundational models are used in software engineering (FM4SE)~\cite{TreudeMSR2023}, we opine that this analysis is just the tip of the iceberg in our quest to make FM4SE fair and equitable to all stakeholders by revealing the latent societal biases and false correlations in foundational models. 
We hypothesise that, as foundational models are trained on large swathes of inherently \textit{human}-generated data, ethical and inclusiveness issues in SE will inevitably creep into such models. From our earlier discussion, these include (but are not limited to): gender bias along the dimensions of SE tasks~\cite{TreudeMSR2023}; differences in pull-request acceptance rates across regions~\cite{RastogiESEM2018,FurtadoIEEESoft2021}; and \textit{intersectional} bias in terms of both gender and region~\cite{PranaTSE2022}.

As the proverbial `canary in the coal mine' --  in the spirit of~\cite{Frermann2023-at} for foundational models powered recruitment and~\cite{TreudeMSR2023} for specifically studying the SE context -- this paper proactively highlights potential concerns of \textit{downstream} applications of GPT (and, by extension, its commercial foundational models counterparts including Google Bard and Meta's Llama) in FM4SE, and proposes future directions for research in this space.

Based on these findings, our future outlook includes a more comprehensive study that scales up the experiment to include more regions of the world, and controlling different attributes to investigate how GPT models (and others in the stable of Foundational Models and LLMs) end up making such decisions to favor one candidate over another.  Therefore, we have proposed six main research agendas (Section \ref{s:researchagenda}) to serve as the foundations for future inquiry along these lines. 

Our short paper clearly demonstrates that these models have inherent biases: In this regard, care should be taken to understand, evaluate, and mitigate such biases and their downstream effects. We hope that this paper serves as a call for action on ethical evaluation, before these models can be deployed `in-the-wild' for FM4SE. 

\section*{Acknowledgment}
This work has been supported by JSPS KAKENHI Nos. JP20H05706, JP23K16862, and JP23K28065.

\bibliographystyle{IEEEtranS.bst}
\bibliography{reference}

\end{document}